\documentclass[12pt, preprint,preprintnumbers,nofootinbib, groupedaddress,superscriptaddress,amsmath,amssymb]{revtex4}
\usepackage{graphicx}% Include figure files
\usepackage{dcolumn}% Align table columns on decimal point
\usepackage{bm}% bold math
\usepackage{amssymb}
\usepackage{amsmath}
\usepackage{epsfig}    
\usepackage{color}
\usepackage{hhline}
\def\be{\begin{equation}}
\def\ee{\end{equation}}
\newcommand{\bea}{\begin{eqnarray}}
\newcommand{\eea}{\end{eqnarray}}
\newcommand{\nn}{\nonumber}

\numberwithin{equation}{section}

\begin{document}
 {\begin{flushright}{CTP-SCU/2021029, APCTP Pre2020 - 020}\end{flushright}}
%%%%%%%%%
\title{Linear seesaw model with a modular $S_4$ flavor symmetry}
% A modular $A_4$ symmetric scotogenic model}
%

\author{Takaaki Nomura}
\email{nomura@scu.edu.cn}
\affiliation{College of Physics, Sichuan University, Chengdu 610065, China}

\author{Hiroshi Okada}
\email{hiroshi.okada@apctp.org}
\affiliation{Asia Pacific Center for Theoretical Physics (APCTP) - Headquarters San 31, Hyoja-dong,
Nam-gu, Pohang 790-784, Korea}
\affiliation{Department of Physics, Pohang University of Science and Technology, Pohang 37673, Republic of Korea}

\date{\today}

\begin{abstract}
We discuss a linear seesaw model with as minimum field content as possible, introducing a modular $S_4$ with the help of gauged $U(1)_{B-L}$ symmetries. Due to rank two neutrino mass matrix, we have a vanishing neutrino mass eigenvalue, and only the normal mass hierarchy of neutrinos is favored through the modular $S_4$ symmetry.
In our numerical $\Delta \chi^2$ analysis, we especially find rather sharp prediction on sum of neutrino masses to be around $60$ meV in addition to the other predictions. 
\end{abstract}
\maketitle
\newpage
\section{Introduction}\label{sec1}
%
\if0
Neutrino sector is theoretically unconfirmed in the standard model (SM), because only two mass squared differences and three mixings are experimentally found.
Thus, a lot of models are proposed  such as canonical seesaw model~\cite{Seesaw1, Seesaw2, Seesaw3, Seesaw4}, inverse seesaw model~\cite{Mohapatra:1986bd, Wyler:1982dd}, linear seesaw model~\cite{Wyler:1982dd, Akhmedov:1995ip, Akhmedov:1995vm}, and so on.
Since these models typically require more free parameters than the other three sectors in the SM fermion, flavor symmetries are also introduced in these models frequently in order to reduce the parameters and get predictions (if possible).
\fi
{
Neutrino sector is theoretically unconfirmed in the standard model (SM), because only two mass squared differences and three mixings are experimentally found and the scale of mass is extremely minuscule compared to the other three sectors in the SM. 
Thus, a lot of scientists expect that neutrino sector would possess new physics.
A gauged $B-L$(baryon number minus lepton number) $U(1)$ symmetry; $U(1)_{B-L}$, is one of the promising prescriptions to generate such tiny neutrino 
masses introducing three right-handed neutrinos with rather heavy masses ($M_R$), that is called canonical seesaw model~\cite{Seesaw1, Seesaw2, Seesaw3, Seesaw4}. Since the mass scale is sometime expected to be the one of
%Planck($M_{Pl}\sim10^{19}$ GeV) or 
a grand unified theory($M_{GUT}\sim10^{15}$ GeV) in order to be small neutrino masses, its scale cannot be confirmed by our current experiments, and the spontaneous $U(1)_{B-L}$ symmetry breaking scale is naturally expected to be the same energy as the cut-off scale; %$M_R\sim M_{Pl}$
 $M_R\sim M_{GUT}$.
 
 In order to achieve a successful neutrino mass model within our scale$(\sim$TeV), another heavy neutral fermions($S_L$) with left-handed chirality are introduced along this line of idea~\footnote{Inverse seesaw model~\cite{Mohapatra:1986bd, Wyler:1982dd} also requests both of $N_R$ and $S_L$ and the neutrino mass could be realized within TeV scale. But this model may not require GUT scale.}. Now that there exist $N_R$ and $S_L$, these can be embedded into the middle scale with $SU(3)_C\otimes SU(2)_L\otimes SU(2)_R \otimes U(1)_{B-L}$~\cite{Wyler:1982dd, Akhmedov:1995ip, Akhmedov:1995vm}, moreover, which is included in $SO(10)$ group~\cite{Malinsky:2005bi}.
Then, the neutrino mass matrix would not be suppressed by the middle scale but by $M_{GUT}$ scale when appropriate charge assignments are assigned for each fields in a supersymmetric theory~\cite{Malinsky:2005bi}. Thus, we can test phenomenologies with our current experiments, supposing the middle scale breaking can occur at TeV scale. This type of model is called by "linear seesaw".
~\footnote{Notice here that our model has a different mechanism from the one of original linear seesaw, even though our neutrino model can be realized within TeV scale. Therefore, our neutrino mass matrix directly depends on the scale of $U(1)_{B-L}$ breaking and ratio between vacuum expectation values (VEVs) of two Higgs doublet model as can be seen in the main text, while we expect higher energy scale such as GUT in order that the modular field must break and get VEV denoted by $\tau$ in our literature. }
%
 % linear seesaw model~\cite{Wyler:1982dd, Akhmedov:1995ip, Akhmedov:1995vm}, and so on.
Since these models typically require more free parameters than the other three sectors in the SM fermion, flavor symmetries are also introduced in these models frequently in order to reduce the parameters and get predictions (if possible).
}

In 2017, attractive flavor symmetries are proposed by papers~\cite{Feruglio:2017spp,
deAdelhartToorop:2011re}, in which they applied modular non-Abelian discrete flavor
symmetries to quark and lepton sectors.
One remarkable advantage is that any dimensionless couplings can also be transformed as non-trivial representations under those symmetries. Therefore, we do not need so many scalars to find a predictive mass matrix. Another advantage is that we have a modular weight from the modular origin that can play a role in stabilizing DM when appropriate charge assignments are distributed to each of the fields of models.
%%%%%  定期分(start) %%%%%%% 
Along the line of this idea, a vast reference has recently appeared in the literature, {\it e.g.},  $A_4$~\cite{Feruglio:2017spp, Criado:2018thu, Kobayashi:2018scp, Okada:2018yrn, Nomura:2019jxj, Okada:2019uoy, deAnda:2018ecu, Novichkov:2018yse, Nomura:2019yft, Okada:2019mjf,Ding:2019zxk, Nomura:2019lnr,Kobayashi:2019xvz,Asaka:2019vev,Zhang:2019ngf, Gui-JunDing:2019wap,Kobayashi:2019gtp,Nomura:2019xsb, Wang:2019xbo,Okada:2020dmb,Okada:2020rjb, Behera:2020lpd, Behera:2020sfe, Nomura:2020opk, Nomura:2020cog, Asaka:2020tmo, Okada:2020ukr, Nagao:2020snm, Okada:2020brs, Yao:2020qyy, Chen:2021zty, Kashav:2021zir, Okada:2021qdf, deMedeirosVarzielas:2021pug, Nomura:2021yjb, Hutauruk:2020xtk, Ding:2021eva, Nagao:2021rio},
%%%
$S_3$ \cite{Kobayashi:2018vbk, Kobayashi:2018wkl, Kobayashi:2019rzp, Okada:2019xqk, Mishra:2020gxg, Du:2020ylx},
%%%
$S_4$ \cite{Penedo:2018nmg, Novichkov:2018ovf, Kobayashi:2019mna, King:2019vhv, Okada:2019lzv, Criado:2019tzk,
Wang:2019ovr, Zhao:2021jxg, King:2021fhl, Ding:2021zbg, Zhang:2021olk, gui-jun},
%%%
$A_5$~\cite{Novichkov:2018nkm, Ding:2019xna,Criado:2019tzk}, double covering of $A_5$~\cite{Wang:2020lxk, Yao:2020zml, Wang:2021mkw, Behera:2021eut}, larger groups~\cite{Baur:2019kwi}, multiple modular symmetries~\cite{deMedeirosVarzielas:2019cyj}, and double covering of $A_4$~\cite{Liu:2019khw, Chen:2020udk, Li:2021buv}, $S_4$~\cite{Novichkov:2020eep, Liu:2020akv}, and the other types of groups \cite{Kikuchi:2020nxn, Almumin:2021fbk, Ding:2021iqp, Feruglio:2021dte, Kikuchi:2021ogn, Novichkov:2021evw} in which masses, mixing, and CP phases for the quark and/or lepton have been predicted~\footnote{For interested readers, we provide some literature reviews, which are useful to understand the non-Abelian group and its applications to flavor structure~\cite{Altarelli:2010gt, Ishimori:2010au, Ishimori:2012zz, Hernandez:2012ra, King:2013eh, King:2014nza, King:2017guk, Petcov:2017ggy}.}.
Moreover, a systematic approach to understanding the origin of CP transformations has been discussed in Ref.~\cite{Baur:2019iai}, 
and CP/flavor violation in models with modular symmetry was discussed in Refs.~\cite{Kobayashi:2019uyt,Novichkov:2019sqv}, 
and a possible correction from K\"ahler potential was discussed in Ref.~\cite{Chen:2019ewa}. Furthermore,
systematic analysis of the fixed points (stabilizers) has been discussed in Ref.~\cite{deMedeirosVarzielas:2020kji}.
{It would be interesting to consider linear seesaw model with local $U(1)_{B-L}(\subset SO(10))$ under modular symmetry since these symmetries can be originated from a string theory. 
Moreover nature of modular symmetry can be used to realize linear seesaw mechanism in addition to constraining flavor structure. }
%%%%%  定期分(end) %%%%%%% 

In this study, we propose a linear seesaw model under modular $S_4$  with the help of $U(1)_{B-L}$ symmetry, in which we try to construct the predictive model as minimum as possible.
Due to rank two neutrino mass matrix, we have a vanishing neutrino mass eigenvalue. Furthermore, only the normal mass hierarchy of neutrinos is favored through the modular $S_4$ symmetry.
In our numerical, we perform $\Delta \chi^2$ analysis in the neutrino sector, considering non-unitarity constraint.

This paper is organized as follows. In Sec.~\ref{sec:model}, we review our model, constructing renormalizable Lagrangian and mass matrices in lepton sector. Then, we formulate the neutrino mass matrix with rank two, in which we estimate structure of the neutrino mass matrix in expansion of modulus. And we derive several observables in lepton sector. At the end of this section we discuss the non-unitarity bound.
 In Sec.~\ref{sec:numerical} we perform $\Delta \chi^2$ analysis in the lepton sector, and show some predictions through our model. In Sec.~\ref{sec:conclusion}, we give summary and discussion.
 In Appendix, we explain the modular $S_4$ symmetry.
 %

%\newpage

\section{Model}
\label{sec:model}
%
% \begin{widetext}
\begin{center} 
\begin{table}[t!]%[tbc]
%\begin{tiny}
\begin{tabular}{|c||c|c|c|c||c|c|c|c|}\hline\hline  
  & \multicolumn{4}{c||}{Fermions} & \multicolumn{3}{c|}{Scalars} \\ \hline \hline
& ~$L_L$~& ~$\overline{e_R},[\overline{\mu_R},\overline{\tau_R}]$~& ~$\overline{N_{R}}$~& ~$S_{L_1},S_{L_2}$~& ~$H_1$~& ~$H_2$~& ~$\varphi$~ \\ \hline \hline 
%%%
$SU(2)_L$ & $\bm{2}$  & $\bm{1}$  & $\bm{1}$  & $\bm{1}$ & $\bm{2}$   & $\bm{2}$& $\bm{1}$     \\\hline 
$U(1)_Y$    & $-\frac12$  & $1$ & $0$  & $0$  & $\frac12$ & $\frac12$& $0$  \\\hline
$U(1)_{B-L}$     & $-1$  & $1$  & $1$  & $0$ & $0$  & $1$ & $-1$   \\\hline
$S_4$   & ${\bf 3'}$ & ${\bf 1}, {\bf 2}$ & $\bm{3}$ & $\bm{1},\bm{1'}$ 
& $\bm{1}$& $\bm{1}$& $\bm{1}$ \\ \hline
$-k_I$  & $-1$ & $-1,-3$ & $-3$ & $-1$ & $0$ & $0$ & $0$ \\
\hline
\end{tabular}
\caption{Lepton and boson particle contents and their charge assignments under $SU(2)_L\times U(1)_Y\times U(1)_{B-L}\times  S_4\times (-k_I)$ where  $L_L\equiv [L_{L_e},L_{L_\mu}, L_{L_\tau}]^T$ $k_I$ is the number of modular weight.}
\label{tab:fields-linear}
% \end{tiny}
\end{table}
\end{center}
%\end{widetext}

\subsection{Model review}

In this section we review our model framework for linear seesaw mechanism, introducing $B-L$ local Abelian symmetry; $U(1)_{B-L}$, and modular $A_4$ symmetry. 
As for fermion sector, we add two left-handed neutral fermions $\{S_{L_1},S_{L_2} \}$ that belong to isospin singlet, where they have zero charge under $U(1)_{B-L}$,  $\{\bm{1},\bm{1'}\}$ under $S_4$, and $-1$ under $-k_I$, respectively.  
Also, we introduce three right-handed neutral fermions $\overline{N_{R}}$ that belong to isospin singlet, where they have $1$ charge under $U(1)_{B-L}$, $\bm{3}$ under $S_4$, and $-3$ under $-k$, respectively.  
%%%
The SM left-handed leptons $L_L\equiv [L_{L_e},L_{L_\mu}, L_{L_\tau}]^T$ belong to $-1$ charge under $U(1)_{B-L}$, $\bm{3'}$ under $S_4$, and $-1$ under $-k$, respectively.  
While the SM right-handed leptons $\{ \overline{e_R},[\overline{\mu_R},\overline{\tau_R}] \}$ belong to $+1$ charge under $U(1)_{B-L}$, $\{ {\bf 1}, {\bf 2} \}$ under $S_4$, and $\{ -1,-3 \}$ under $-k_I$, respectively.  

As for scalar sector, we adopt two Higgs doublet $H_1, H_2$ plus an isospin singlet field $\varphi$. 
An isospin singlet $\varphi$ has $-1$ charge under the $U(1)_{B-L}$.
Here, $H_1$ is SM-like Higgs that has zero charge under $U(1)_{B-L}$ and $-k_I$
while $H_2$ has $1$ charge under $U(1)_{B-L}$, and zero under $-k_I$.
We denote each of vacuum expectation values (VEVs) to be $\langle H_{1,2} \rangle\equiv [0,v_{1,2}/\sqrt2]^T$, and $\langle \varphi \rangle\equiv v_{\varphi} /\sqrt2$.
We summarize our particle content and their assignments in Table \ref{tab:fields-linear}.

Then  the valid lepton Yukawa Lagrangian is symbolized by 
\begin{align}\label{a4lag}
 -\mathcal{L}_{\rm lepton} = \mathcal{L}_{M_\ell} + \mathcal{L}_{\rm M_D}+\mathcal{L}_{\rm M'_D}
 +\mathcal{L}_{M_{NS}}  ,
\end{align}
where $\mathcal{L}_{M_\ell} $ is charged lepton Yukawa Lagrangian.
$\mathcal{L}_{\rm M_D}$ is the one of [$\overline{N_R} L_L \tilde H_1$] where $\tilde H\equiv i\sigma_2 H^*$.
$\mathcal{L}_{\rm M'_D}$ is the one of [$\overline{L_L^C} S_L  H_2$].
$\mathcal{L}_{M_{NS}} $ is the one of [$\overline{N_R} S_L \varphi$].
$[\cdots]$ implies that concrete flavor structures are manifolded.
We will see each of their structures below.

{The scalar potential of our model is written by
\begin{align}
V & = m_\varphi^2 \varphi^*\varphi + m_1^2 H_1^\dagger H_1 + m_2^2 H_2^\dagger H_2 - \mu_{12} (H_1^\dagger H_2 \varphi + h.c.) \nonumber \\
& + \lambda_1 (H_1^\dagger H_1)^2+ \lambda_2 (H_2^\dagger H_2)^2 \nonumber + \lambda_\varphi (\varphi^* \varphi)^2  +  \lambda_3 (H_1^\dagger H_1)(H_2^\dagger H_2) + \lambda_4 (H_1^\dagger H_2) (H_2^\dagger H_1) 
\nonumber \\
& + \lambda_{\varphi H_1} (H_1^\dagger H_1) (\varphi^* \varphi)+ \lambda_{\varphi H_2} (H_2^\dagger H_2) (\varphi^* \varphi),
\end{align}
where $h.c.$ stands for Hermitian conjugate.
We consider $\varphi$ develops a VEV at much higher scale than $H_{1,2}$. 
Then after $\varphi$ developing a VEV the scalar potential matches with that of two Higgs doublet model (THDM) without $(H_1^\dagger H_2)^2$ term due to $U(1)_{B-L}$ symmetry.
Also the Yukawa couplings associated with two Higgs doublet are those of type-I THDM since only $H_1$ can couples to the SM fermions.
In our analysis below we thus do not discuss THDM part further and focus on neutrino mass.}

%{\color{red}
\subsection{Valid Lagrangians}
Before discussing the valid Lagrangians for lepton sector, we define Yukawa couplings under modular $S_4$ symmetry as follows: $S_4$ doublet with $-k_I=2$ and $4$ are respectively denoted by $Y^{(2)}_{\bf 2}\equiv [y_1,y_2]^T$ and $[y'_1,y'_2]^T$, $S_4$ triplet with $-k_I=2$ is $Y^{(2)}_{\bf 3'}\equiv [y_3,y_4,y_5]^T$, 
$S_4$ triplets with $-k_I=4$ are $Y^{(4)}_{\bf 3}\equiv [y'_3,y'_4,y'_5]^T$, and $Y^{(4)}_{\bf 3'}\equiv [y''_3,y''_4,y''_5]^T$. Each of structures is explicitly written in Appendix.

%\noindent \\
{\bf \underline{Charged lepton mass matrix}}: \\
The renormalizable Lagrangian for charged-lepton sector is given by 
\begin{align}
-\mathcal{L}_{M_\ell} &=
\alpha_\ell \overline{e_R}
(y_3 L_{L_e} +y_4 \bar L_{L_\tau} +y_5 \bar L_{L_\mu}) \tilde H_1 \nn\\
&+ \beta_\ell\left[\frac{\sqrt3}{2}\overline{\mu_R}(y'_4 L_{L_\mu} +y'_5 L_{L_\tau}) 
+\overline{\tau_{R}} \left(-y'_3 L_{L_e} + \frac12 (y'_4 L_{L_\tau} +y'_5 L_{L_\mu}) \right) \right] \tilde H_1 \nn \\
&+ \gamma_\ell\left[\frac{\sqrt3}{2}\overline{\tau_R}(y''_4 L_{L_\mu} +y''_5 L_{L_\tau}) 
+\overline{\mu_{R}} \left(y''_3 L_{L_e} - \frac12 (y''_4 L_{L_\tau} +y''_5 L_{L_\mu}) \right) \right] \tilde H_1 
+{\rm h.c.},
\end{align}
where $\{\alpha_\ell,\beta_\ell,\gamma_\ell \}$ are real parameters without loss of generality.
Then the charged-lepton mass matrix after the spontaneous symmetry breaking is given by
\begin{align}
(M_\ell)_{RL} =\frac{v_1}{\sqrt2}
 \begin{pmatrix}  \alpha_\ell y_3  &  \alpha_\ell y_5 &  \alpha_\ell y_4 \\
\gamma_\ell y''_3  & \frac{\sqrt3}{2}\beta_\ell y'_4-\frac12 \gamma_\ell y''_5 
& \frac{\sqrt3}{2}\beta_\ell y'_5-\frac12 \gamma_\ell y''_4 \\
-\beta_\ell y'_3  & \frac{\sqrt3}{2}\gamma_\ell y''_4+\frac12 \beta_\ell y'_5 
& \frac{\sqrt3}{2}\gamma_\ell y''_5+\frac12 \beta_\ell y'_4 \\ \end{pmatrix}             
\label{Eq:Mell} .
\end{align}
The charged-lepton mass eigenvalues are obtained by diagonalizing ${\rm diag}[m_e,m_\mu,m_\tau] = V_{R_\ell}^\dag M_\ell V_{L_\ell}$, where $V_{L_\ell, R_\ell}$ are unitary matrices. 
%%%
In our numerical analysis, we will determine the free parameters {$\{\alpha_\ell,\beta_\ell,\gamma_\ell\}$} so as to fit the three charged-lepton mass eigenstates after giving all the numerical values, by applying the relations:
%%%
Here, we fix $\alpha_\ell,\beta_\ell,\gamma_\ell$ in order to find the experimental three charged-lepton masses,
applying the following relations: 
\begin{align}
&{\rm Tr}[{M_\ell}^\dag M_\ell] = |m_e|^2 + |m_\mu|^2 + |m_\tau|^2,\quad
 {\rm Det}[{M_\ell}^\dag M_\ell] = |m_e|^2  |m_\mu|^2  |m_\tau|^2,\nn\\
&({\rm Tr}[{M_\ell}^\dag] M_\ell)^2 -{\rm Tr}[({M_\ell}^\dag M_\ell)^2] =2( |m_e|^2  |m_\mu|^2 + |m_\mu|^2  |m_\tau|^2+ |m_e|^2  |m_\tau|^2 ).\label{eq:l-cond}
\end{align}

\noindent
{\bf \underline{Neutral fermion mass matrices}:}\\
At first, we construct the valid Lagrangian of Dirac mass matrix $\mathcal{L}_{\rm M_D}$.
This is given by 
\begin{align}
\mathcal{L}_{\rm M_D} 
&= 
\alpha_D
\left[y'_3(\overline{N_{R_3}} L_{L_\mu}-\overline{N_{R_2}} L_{L_\tau})  
+
y'_4(\overline{N_{R_1}} L_{L_\tau}-\overline{N_{R_3}} L_{L_e})  
+
y'_5(\overline{N_{R_2}} L_{L_e}-\overline{N_{R_1}} L_{L_\mu})\right] \tilde H_1 \nn\\
&+
\beta_D
\left[y''_3(\overline{N_{R_3}} L_{L_\tau}-\overline{N_{R_2}} L_{L_\mu})  
+
y''_4(-\overline{N_{R_1}} L_{L_\mu}-\overline{N_{R_2}} L_{L_e})  
+
y''_5(\overline{N_{R_1}} L_{L_\tau}+\overline{N_{R_3}} L_{L_e})\right] \tilde H_1 \nn\\
&+
\gamma_D
\left[\frac{\sqrt3}{2}y'_1(\overline{N_{R_2}} L_{L_\mu}+\overline{N_{R_3}} L_{L_\tau})  
+
y'_2
\{-\overline{N_{R_1}} L_{L_e}
+\frac12
(\overline{N_{R_2}} L_{L_\tau}+\overline{N_{R_3}} L_{L_\mu})\}\right] \tilde H_1+{\rm h.c.},
\end{align}
where we suppose $\alpha_D$ to be real while $\beta_D,\ \gamma_D$ to be complex after rephasing of fields.
%%%
Similar to the charged-lepton sector we find the Dirac mass matrix as follows:
\begin{align}
(M_D)_{RL}  = 
\frac{v_1}{\sqrt2}
 \begin{pmatrix}  -\gamma_D y'_2  &  -\alpha_D y'_5 - \beta_D y''_4&  \alpha_D y'_4+\beta_D y''_5 \\
\alpha_D y'_5 - \beta_D y''_4 &-\beta_D y''_3+ \frac{\sqrt3}{2}\gamma_D y'_1
&-\alpha_D y'_3 + \frac12 \gamma_D y'_2 \\
- \alpha_D y'_4+\beta_D y''_5  &\alpha_D y'_3 + \frac12 \gamma_D y'_2
& \beta_D y''_3+ \frac{\sqrt3}{2}\gamma_D y'_1 \\ \end{pmatrix}     
\label{Eq:md} .
\end{align}
For our convenience to analyze the neutrino oscillation, we redefine $M_D\equiv \frac{v_1}{\sqrt2}\tilde{M_D}$ 

%%%%%%%%%%%%%%%%%%%%%%

Another Dirac Lagrangian is induced via $\mathcal{L}_{\rm M'_D}$, which is given by 
\begin{align}
\mathcal{L}_{\rm M'_D} 
&= 
\alpha'_D
\left[y_3 \overline{L^C_{L_e}} +y_4 \overline{L^C_{L_\tau}}+y_5 \overline{L^C_{L_\mu}}\right]S_{L_1} H_2
+{\rm h.c.},
\end{align}
%%%
Then, we find another Dirac mass matrix as follows:
\begin{align}
(M'_D)_{L_L S_L}  = 
\frac{v_2}{\sqrt2} \alpha'_D
 \begin{pmatrix} 
  y_3  &  0 \\
  y_5  &  0 \\
y_4  &  0 \\
 \end{pmatrix}     
\label{Eq:md} .
\end{align}
Being the same as the reason for $M_D$, we redefine $M'_D\equiv\frac{v_2}{\sqrt2} \alpha'_D \tilde{M'_D}$ .

%%%%%%%%%%%%%%%%%%%%%%

The third term of Lagrangian $\mathcal{L}_{M_{NS}} $ is given by
\begin{align}
\mathcal{L}_{M_{NS}} = &
\alpha_{NS}
\left[y'_3 \overline{N_{R_e}} + y'_4 \overline{N_{R_\tau}} + y'_5\overline{N_{R_\mu}}\right] S_{L_1}\varphi
+
\beta_{NS}
\left[y''_3 \overline{N_{R_e}} + y''_4 \overline{N_{R_\tau}} + y''_5\overline{N_{R_\mu}}\right] S_{L_2}\varphi
+{\rm h.c.},
\end{align}
where $\alpha_{NS}, \beta_{NS}$ are real without loss of generality.
%%%
Then, we obtain the mass matrix
\begin{align}
M_{NS} =\frac{v_\varphi}{\sqrt2}
%%%
\begin{pmatrix}  
y'_3  &  y''_3  \\
y'_5  &  y''_5  \\
y'_4  &  y''_4  \\
\end{pmatrix}  
%%%               
\begin{pmatrix}  
\alpha_{NS}  & 0  \\
0  & \beta_{NS}  \\
\end{pmatrix}  
\equiv
\frac{v_\varphi}{\sqrt2} \tilde M_{N S}
\label{Eq:Mell} .
\end{align}

In basis of $[\nu_L,N_R^C,S_L]^T$, the neutral fermion mass matrix is given by
\begin{align}
M_N=
 \begin{pmatrix}  0_{3\times3}  &  M_D^T &  M'_D \\
M_D  & 0_{3\times3} & M_{NS} \\
m'^T_D  &  M_{NS}^T  & 0_{2\times2}
\end{pmatrix}   
\label{Eq:neut} .
\end{align}
Then, the active neutrino mass matrix is given by
\begin{align}
m_\nu
&= 
M'_D (M_{NS}^T M_{NS})^{-1}M_{NS}^T M_D + [M'_D (M_{NS}^T M_{NS})^{-1}M_{NS}^T M_D]^T\nn\\
& =
\frac{v_1v_2}{\sqrt{2} v_\varphi}
\left(
\tilde M'_D (\tilde M_{NS}^T \tilde M_{NS})^{-1}\tilde M_{NS}^T\tilde M_D + [\tilde M'_D (\tilde M_{NS}^T\tilde M_{NS})^{-1}\tilde M_{NS}^T \tilde M_D]^T
\right)\nn\\
%%%
&=%\frac{v_1v_2}{\sqrt{2} v_\varphi}
\kappa \tilde m_\nu, 
\label{Eq:act-neut} 
\end{align}
where $\kappa\equiv \frac{v_1v_2}{\sqrt{2} v_\varphi}$ and we assume mass hierarchies among $M_D, M'_D \ll M_{NS}$.~\footnote{Mass hierarchies is dynamically achieved in refs.~\cite{Wang:2015saa, Das:2017ski}.}.
The neutrino mass eigenvalues are obtained as follows:
$D_\nu=\kappa D_\nu= U_\nu^T m_\nu U_\nu=\kappa U_\nu^T \tilde m_\nu U_\nu$, where $U_\nu$ is a unitary matrix. Then, the Pontecorvo-Maki-Nakagawa-Sakata (PMNS) matrix is given by $U_{PMNS}\equiv V^\dag_{L_\ell} U_\nu$. Notice that $m_\nu$ is rank two, thus the lightest neutrino mass is zero. 
% since the charged-lepton is not diagonal basis in original Lagrangian.
Here $\kappa$ is described by one experimental values and dimensionless neutrino mass eigenstates as follows:
\begin{align}
{\rm (NH)}:\  \kappa^2= \frac{|\Delta m_{\rm atm}^2|}{\tilde D_{\nu_3}^2},
\quad
{\rm (IH)}:\  \kappa^2= \frac{|\Delta m_{\rm atm}^2|}{\tilde D_{\nu_2}^2},
 \end{align}
where $\Delta m_{\rm atm}^2$ is the atmospheric neutrino mass-squared difference and NH(IH) stands for normal(inverted) ordering respectively. 
Subsequently, the solar mass difference squared can be written in terms of $\kappa$ as follows:
\begin{align}
{\rm (NH)}:\  \Delta m_{\rm sol}^2= {\kappa^2} {\tilde D_{\nu_2}^2},
\quad
{\rm (IH)}:\  \Delta m_{\rm sol}^2= {\kappa^2}({\tilde D_{\nu_2}^2-\tilde D_{\nu_1}^2}),
 \end{align}
 which can be compared to the observed value.
 In other words, we explicitly write  the mass eigenvalues in terms of $\Delta m_{\rm atm}^2$ and $\Delta m_{\rm sol}^2$ as follows:
 \begin{align}
&{\rm (NH)}:\ D_{\nu_1}^2=0, \ D_{\nu_2}^2=\Delta m_{\rm sol}^2, \ D_{\nu_3}^2=\Delta m_{\rm atm}^2,\\
&{\rm (IH)}:\ D_{\nu_1}^2=\Delta m_{\rm atm}^2 - \Delta m_{\rm sol}^2, \ D_{\nu_2}^2=\Delta m_{\rm sol}^2, \ D_{\nu_3}^2=0,
 \end{align}
 which implies that NH is hierarchical but IH is degenerate, since $ \Delta m_{\rm sol}^2/ \Delta m_{\rm atm}^2<<1$.
Here, we expand $|\tilde m_\nu|^2$ in terms of $q\equiv e^{2p\pi i\tau}<<1$.
Then the mass matrix is given by 
\begin{align}
|\tilde m_\nu|^2\sim
\begin{pmatrix} 
 {\cal O}(1) & {\cal O}(q)  & 0    \\ 
{\cal O}(q) &   {\cal O}(1)  &  {\cal O}(q^2) \\
0 &  {\cal O}(q^2) & 0
\end{pmatrix}.
\end{align}
The ratio between two nonzero squared eigenvalues $R$ is estimated by
\begin{align}
R =  {\cal O}(q) <<1 .
\end{align}
It suggests that the neutrino mass eigenvalues tend to be hierarchical, therefore NH is favored.
In fact, we would not obtain the allowed region within $3\sigma$ for IH in our numerical analysis.
Thus, we focus on NH hereafter.
 
%Tr$[D_{\nu}] \lesssim$ 0.12 eV is given by the recent cosmological data~\cite{Aghanim:2018eyx, Vagnozzi:2017ovm}.
{ In our model, PMNS matrix is parametrized by three mixing angle $\theta_{ij} (i,j=1,2,3; i < j)$, one CP violating Dirac phase $\delta_{CP}$,
and one Majorana phase $\alpha_{21}$ as follows:
\begin{equation}
U_{PMNS} = 
\begin{pmatrix} c_{12} c_{13} & s_{12} c_{13} & s_{13} e^{-i \delta_{CP}} \\ 
-s_{12} c_{23} - c_{12} s_{23} s_{13} e^{i \delta_{CP}} & c_{12} c_{23} - s_{12} s_{23} s_{13} e^{i \delta_{CP}} & s_{23} c_{13} \\
s_{12} s_{23} - c_{12} c_{23} s_{13} e^{i \delta_{CP}} & -c_{12} s_{23} - s_{12} c_{23} s_{13} e^{i \delta_{CP}} & c_{23} c_{13} 
\end{pmatrix}
\begin{pmatrix} 1 & 0 & 0 \\ 0 & e^{i \frac{\alpha_{21}}{2}} & 0 \\ 0 & 0 & 1 \end{pmatrix},
\end{equation}
where $c_{ij}$ and $s_{ij}$ stand for $\cos \theta_{ij}$ and $\sin \theta_{ij}$ respectively. 
Then, these mixings are given in terms of the components of $U_{PMNS}$ as follows:
\begin{align}
\sin^2\theta_{13}=|(U_{PMNS})_{13}|^2,\quad 
\sin^2\theta_{23}=\frac{|(U_{PMNS})_{23}|^2}{1-|(U_{PMNS})_{13}|^2},\quad 
\sin^2\theta_{12}=\frac{|(U_{PMNS})_{12}|^2}{1-|(U_{PMNS})_{13}|^2}.
\end{align}
Also we compute the Jarlskog invariant
 $J_{CP}$ that is derived from PMNS matrix elements as follows:
%$\delta_{CP}$ derived from PMNS matrix elements $U_{\alpha i}$ such that
\begin{equation}
J_{CP} = \text{Im} [U_{e1} U_{\mu 2} U_{e 2}^* U_{\mu 1}^*] = s_{23} c_{23} s_{12} c_{12} s_{13} c^2_{13} \sin \delta_{CP}.
\end{equation}
Majorana phase is estimated in terms of other invariant $I_1$ as follows:
\begin{equation}
I_1 = \text{Im}[U^*_{e1} U_{e2}] = c_{12} s_{12} c_{13}^2 \sin \left( \frac{\alpha_{21}}{2} \right).
\end{equation}
In addition, the effective mass for the neutrinoless double beta decay is written by
\begin{align}
&{\rm (NH)}:\ \langle m_{ee}\rangle=\kappa|\tilde D_{\nu_2} s^2_{12} c^2_{13}e^{i\alpha_{21}}+\tilde D_{\nu_3} s^2_{13}e^{-2i\delta_{CP}}|,\\
&{\rm (IH)}:\ \langle m_{ee}\rangle=\kappa|\tilde D_{\nu_1} c^2_{12} c^2_{13}+\tilde D_{\nu_2} s^2_{12} c^2_{13}e^{i\alpha_{21}}|,
\end{align}
where its value could be measured by KamLAND-Zen in future~\cite{KamLAND-Zen:2016pfg}. 
We will adopt the neutrino experimental data at 3$\sigma$ interval in Nufit 5.0~\cite{Esteban:2018azc,Nufit}.
\if0
as follows:
\begin{align}
&{\rm NO}: \Delta m^2_{\rm atm}=[2.432, 2.618]\times 10^{-3}\ {\rm eV}^2,\
\Delta m^2_{\rm sol}=[6.79, 8.01]\times 10^{-5}\ {\rm eV}^2,\\
&\sin^2\theta_{13}=[0.02046, 0.02440],\ 
\sin^2\theta_{23}=[0.427, 0.609],\ 
\sin^2\theta_{12}=[0.275, 0.350],\nn\\
%%%
&{\rm IO}: \Delta m^2_{\rm atm}=[2.416, 2.603]\times 10^{-3}\ {\rm eV}^2,\
\Delta m^2_{\rm sol}=[6.79, 8.01]\times 10^{-5}\ {\rm eV}^2,\\
&\sin^2\theta_{13}=[0.02066, 0.02461],\ 
\sin^2\theta_{23}=[0.430, 0.612],\ 
\sin^2\theta_{12}=[0.275, 0.350].\nn
\end{align}
\fi

\noindent
{\bf \underline{Non-unitarity}}: \\
Here, let us briefly discuss non-unitarity matrix $U'_{PMNS}$.
This is typically parametrized by the form 
\begin{align}
U'_{PMNS}\equiv \left(1-\frac12 F^\dag F\right) U_{PMNS},
\end{align}
where $F\equiv  (M_{NS}^T M_{NS})^{-1}M_{NS}^T M_D$ is a hermitian matrix, and $U'_{PMNS}$ represents the deviation from the unitarity. 
%
%The global constraints are found via several experimental results such as the SM $W$ boson mass $M_W$, the effective Weinberg angle $\theta_W$, several ratios of $Z$ boson fermionic decays, invisible decay of $Z$, electroweak universality, measured Cabbibo-Kobayashi-Maskawa, and lepton flavor violations
Applying global constraints~\cite{Fernandez-Martinez:2016lgt},
one finds~\cite{Agostinho:2017wfs}
\begin{align}
|FF^\dag|\le  
\left[\begin{array}{ccc} 
2.5\times 10^{-3} & 2.4\times 10^{-5}  & 2.7\times 10^{-3}  \\
2.4\times 10^{-5}  & 4.0\times 10^{-4}  & 1.2\times 10^{-3}  \\
2.7\times 10^{-3}  & 1.2\times 10^{-3}  & 5.6\times 10^{-3} \\
 \end{array}\right].
\end{align} 
%%%
In our case, $F\equiv    (M_{NS}^T M_{NS})^{-1}M_{NS}^T M_D=\frac{v_1}{v_\varphi}
(\tilde M_{NS}^T\tilde M_{NS})^{-1}\tilde M_{NS}^T \tilde M_D$.
Since  $v_\varphi$ is freely taken to be large (while $v_1={\cal O}(100)$ GeV at most), we easily satisfy this bound.
% we obtain $(v_1/v_\varphi)^2\approx 10^{-3}$, therefore we demand $(\tilde M_{NS}^T\tilde M_{NS})^{-1}\tilde M_{NS}^T \tilde M_D\approx {\cal O}(10^{-2})$.

 \section{Numerical analysis}
\label{sec:numerical}
In this section, we carry out numerical $\Delta \chi^2$ analysis searching for parameters satisfying neutrino oscillation data and non-unitarity constraint, and show our predictions, where we use the best fit values of charged-lepton masses.
Notice here that we focus on NH since IH is disfavored by analytical estimation as can be seen in the previous section.

In our numerical analysis, we randomly scan free parameters in following ranges 
\begin{align}
%&|{\rm Re}[\tau]| \in [0,0.5],\quad {\rm Im}[\tau]\in [0.5,2], \nonumber \\
& \{ \alpha_D, |\beta_D|, |\gamma_D|, \alpha_{NS}, \beta_{NS} \} \in [10^{-5},1.0],
\quad t_\beta\in[10,100],
\end{align}
where $\tau$ runs over the fundamental region, and $t_\beta\equiv v_1/v_2$ and $\sqrt{v_1^2+v_2^2}=246$ GeV.
Under the above regions, we perform numerical analysis.
Fig.~\ref{fig:1} shows the correlation between real part of $\tau$ and imaginary part of $\tau$, where the blue points are allowed within 2, green ones within 3, and red one within 5 of $\Delta \chi^2$ analysis
for five accurately known observables $\Delta m^2_{\rm atm},\Delta m^2_{\rm sol},s_{12}^2,s_{23}^2,s_{13}^2$ in Nufit 5.0~\cite{Esteban:2018azc,Nufit}.
%\UTF{2206}m2atm, \UTF{2206}m2sol, sin2 θ12, sin2 θ23 and sin2 θ13 in NuFit 5.0 [133]..
The real part runs whole the range, but the imaginary part is localized at the region of $[1.35-1.5]$.
%%%%%%%%%%%%%%%%%%%%%%%%%%%%%%%%%%%%%%%%%%%%%%%%%%%%%%%%%%%%%%%%%%%%%%%%%%%%%%%%%%%%
%%%%%%%%%%%%%%%%%%%
\begin{figure}[tb!]\begin{center}
\includegraphics[width=80mm]{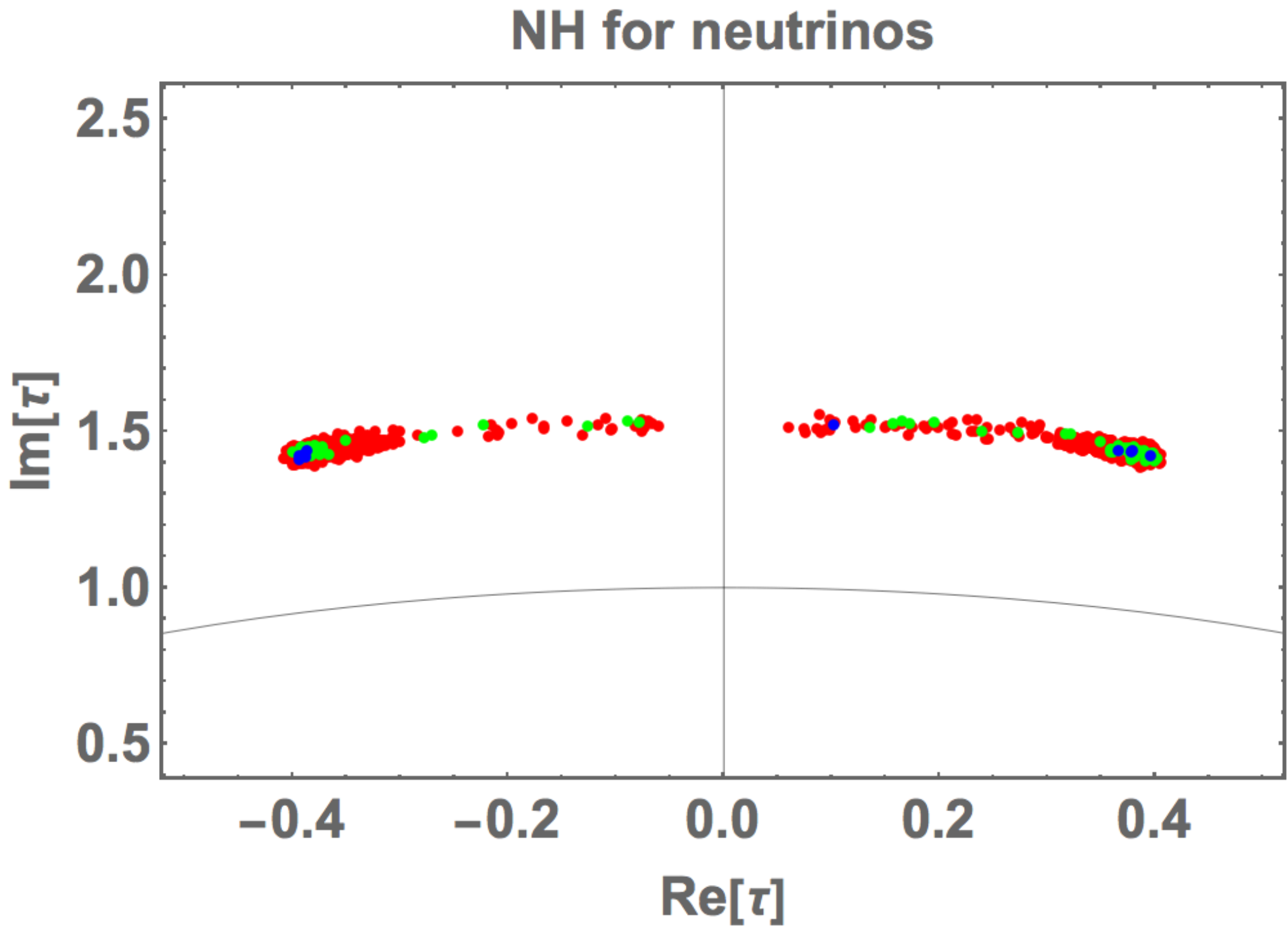} 
\caption{The allowed region of modulus $\tau$ , where the blue points are allowed within 2, green ones within 3, and red one within 5 of $\Delta \chi^2$ analysis.}   
\label{fig:1}\end{center}\end{figure}
%%%%%%%%%%%%%%%%%%%
%%%%%%%%%%%%%%%%%%%%%%%%%%%%%%%%%%%%%%%%%%%%%%%%%%%%%%%%%%%%%%%%%%%%%%%%%%%%%%%%%%%%

Fig.~\ref{fig:2} demonstrates the correlation between the Majorana phase $\alpha_{21}$ and Dirac CP phase $\delta_{CP}$. The legend is the same as the case of Fig.1.
We find a clear feature on the region that $\alpha_{21}$ is within $\{0^\circ - 120^\circ, 240^\circ - 310^\circ \}$,
while $\delta_{CP}$ is within $\{70^\circ - 290^\circ \}$.

%%%%%%%%%%%%%%%%%%%%%%%%%%%%%%%%%%%%%%%%%%%%%%%%%%%%%%%%%%%%%%%%%%%%%%%%%%%%%%%%%%%%
%%%%%%%%%%%%%%%%%%%
\begin{figure}[tb!]\begin{center}
\includegraphics[width=80mm]{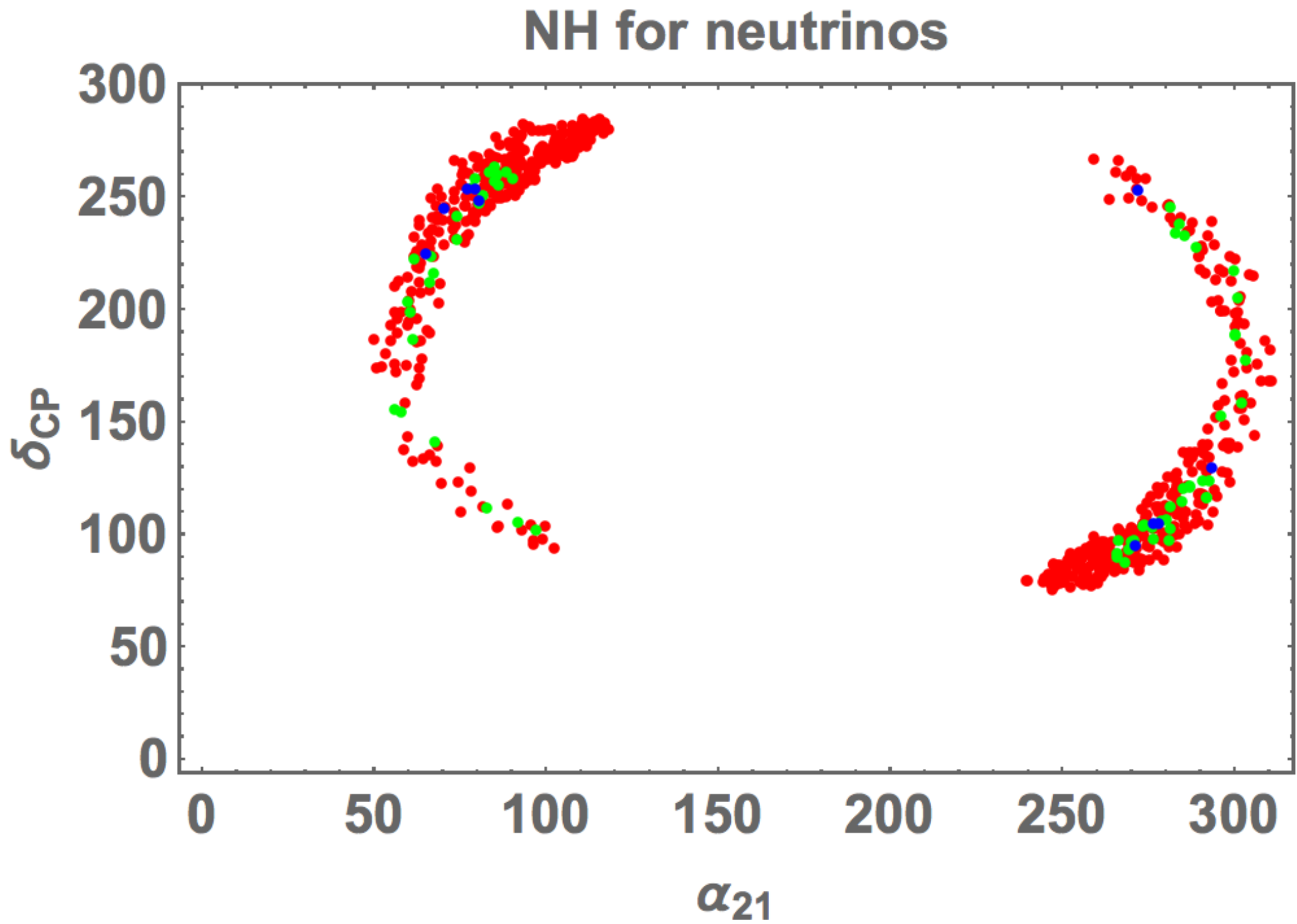}
\caption{Correlation between a Majorana phase $\alpha_{21}$ and $\delta_{CP}$, where the legend is the same as the case of Fig.1. }   
\label{fig:2}\end{center}\end{figure}
%%%%%%%%%%%%%%%%%%%
%%%%%%%%%%%%%%%%%%%%%%%%%%%%%%%%%%%%%%%%%%%%%%%%%%%%%%%%%%%%%%%%%%%%%%%%%%%%%%%%%%%%
%\begin{figure}[tb!]\begin{center}
% \includegraphics[width=80mm]{lsms4-s23-dcp_nh.pdf}
% \caption{Upper figures: Predicted correlation between the effective mass for the neutrinoless double beta decay $\langle m_{ee} \rangle$ and Dirac-CP phase $\delta^\ell_{CP}$. 
% Lower figures: Predicted correlation between $\langle m_{ee} \rangle$ and sum of neutrino mass $\sum m$. The left-(right-)side figures correspond to NO(IO).}   \label{fig:3}\end{center}\end{figure}
%%%%%%%%%%%%%%%%%%%
%%%%%%%%%%%%%%%%%%%%%%%%%%%%%%%%%%%%%%%%%%%%%%%%%%%%%%%%%%%%%%%%%%%%%%%%%%%%%%%%%%%%

The upper figures in Fig.~\ref{fig:3} show correlation between the effective mass for the neutrinoless double beta decay $\langle m_{ee} \rangle$ and the sum of neutrino masses $\sum m_i$ in unit of [eV], where the legend is the same as the case of Fig.1.
We find that $\langle m_{ee} \rangle$ is allowed within $[0.001 - 0.004]$ eV.
On the other hand $\sum m_i$ is restricted to be around $0.06$ eV, which would be a sharp prediction of this model.
%%%%%%%%%%%%%%%%%%%%%%%%%%%%%%%%%%%%%%%%%%%%%%%%%%%%%%%%%%%%%%%%%%%%%%%%%%%%%%%%%%%%
%%%%%%%%%%%%%%%%%%%
\begin{figure}[tb!]\begin{center}
\includegraphics[width=80mm]{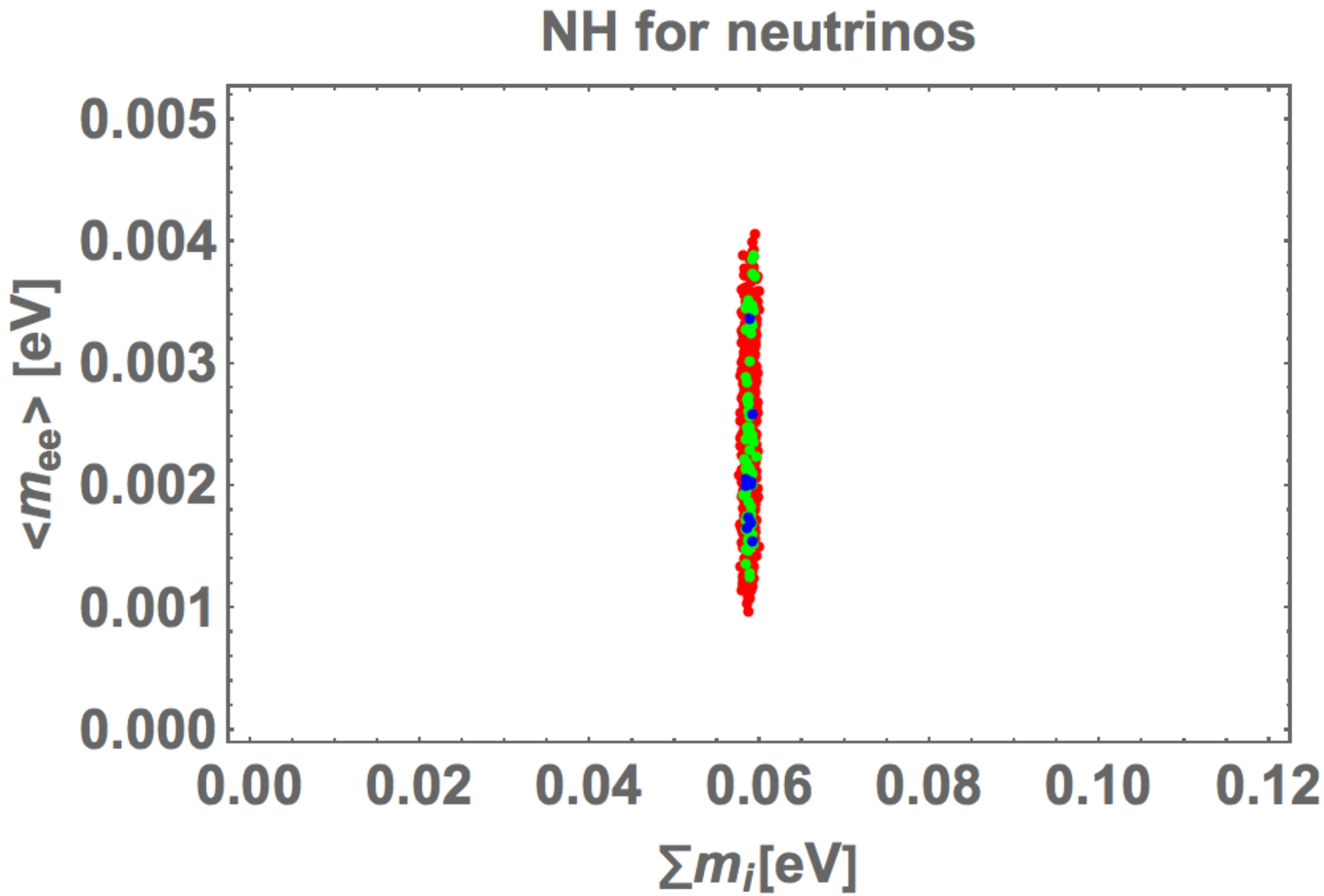}
\caption{Predicted correlation between the effective mass for the neutrinoless double beta decay $\langle m_{ee} \rangle$ and the sum of neutrino masses $\sum m_i$ in unit of [eV], where the legend is the same as the case of Fig.1. }   
\label{fig:3}\end{center}\end{figure}
%%%%%%%%%%%%%%%%%%%
%%%%%%%%%%%%%%%%%%%%%%%%%%%%%%%%%%%%%%%%%%%%%%%%%%%%%%%%%%%%%%%%%%%%%%%%%%%%%%%%%%%%

%
Finally, we show a benchmark in table~\ref{bp-tab_nh}, where we select it so that $\sqrt{\Delta \chi^2}$ is minimum.
The mass matrices for dimensionless neutrino and charged-lepton are found as
\begin{align}
\tilde m_\nu&=
\left[\begin{array}{ccc} 
37.1331- 17.9304 i &  -16.2576 + 55.2032 i  & 35.2235+ 24.3462 i  \\
-16.2576 + 55.2032 i &-40.0223 -38.1947 i  &  -55.9229 + 44.0753 i \\
35.2235+ 24.3462 i  & -55.9229 + 44.0753 i &-1.23231 + 74.8642 i \\
 \end{array}\right],\\
 %%%
 M_\ell&=
\left[\begin{array}{ccc} 
0.00232 + 0.000467 i & -0.00111 + 0.00314 i & -0.00453 +  0.00274 i  \\
-0.282 + 0.102 i & -0.0524 - 0.469 i& 0.214 - 0.559 i \\
0.336 - 0.277 i & 0.362 + 0.706 i& -0.160 + 1.23 i \\
 \end{array}\right].
 \end{align}

%%%%%%%%%%%%%%%%%%%%%%%%%%%%%%%%%%%%%%%%%%%%%%%%%%%
\begin{table}[h]
	\centering
	\begin{tabular}{|c|c|} \hline 
			\rule[14pt]{0pt}{0pt}
		$\tau$ & $0.113762 + 1.43906 i$     \\ \hline
		\rule[14pt]{0pt}{0pt}
		%%%
		$t_\beta$ & $98.6$     \\ \hline
		\rule[14pt]{0pt}{0pt}
		%%%
%		
		$[ \alpha_\ell, \gamma_{\ell}, \beta_{\ell}]$ & $[1.72\times 10^{-5},\ 6.15\times 10^{-4},\ 8.06\times 10^{-4}]$ \\ \hline
		\rule[14pt]{0pt}{0pt}
		$[ \alpha_D, \alpha_{NS}, \beta_{NS}]$ & $[-0.0112907,\ -0.00078512,\ 0.0203379]$ \\ \hline
		\rule[14pt]{0pt}{0pt}
%		$c_\eta$  &  $0.00767509 - 0.0313634 i$ & $0.0013744 - 0.0000470787 I$ & $-0.00227049 + 0.00350111 I$\\
		$[ \beta_D, \gamma_D]$ & $[-0.01432 - 0.00360 i, -1.90\times 10^{-5} - 2.34\times10^{-6} i]$  \\ \hline
		\rule[14pt]{0pt}{0pt}
		%%%
		$\Delta m^2_{\rm atm}$  &  $2.53\times10^{-3} {\rm eV}^2$    \\ \hline
		\rule[14pt]{0pt}{0pt}
%		$c'$  &  $-0.213667 - 0.271705 i$ & $-0.153721 + 0.0313641 I$ & $-3.40695 \times 10^{-6} - 0.0000598624 I$\\
		$\Delta m^2_{\rm sol}$  &  $7.48\times10^{-5} {\rm eV}^2$       \\ \hline
		\rule[14pt]{0pt}{0pt}
%		$\sin^2\theta_{12}$ & $ 0.322231$& $0.28036$ & $0.33157$\\
		$\sin^2\theta_{12}$ & $ 0.289$ \\ \hline
		\rule[14pt]{0pt}{0pt}
%		$\sin^2\theta_{23}$ &  $ 0.563489$& $0.463487$ & $0.579596$\\
		$\sin^2\theta_{23}$ &  $ 0.565$    \\ \hline
		\rule[14pt]{0pt}{0pt}
%		$\sin^2\theta_{13}$ &  $ 0.0235092$&$0.0240532$ & $0.0218055$\\
		$\sin^2\theta_{13}$ &  $ 0.02207$   \\ \hline
		\rule[14pt]{0pt}{0pt}
%		$\delta_{CP}^\ell$ &  $328.932^\circ$& $ 170.523^\circ$ & $335.678^\circ$\\
		$[\delta_{CP},\ \alpha_{21} ]$ &  $[248^\circ,\, 80.3^\circ]$    \\ \hline
		\rule[14pt]{0pt}{0pt}
		$\sum m_i$ &  $58.5$\,meV    \\ \hline
		\rule[14pt]{0pt}{0pt}
		$\langle m_{ee} \rangle$ &  $1.65$\,meV     \\ \hline
		\rule[14pt]{0pt}{0pt}
%		$\sqrt{\chi^2}$ &  $2.86953$ & $4.78971$  & $4.54076$\\
		$\sqrt{\Delta\chi^2}$ &  $1.40$    \\ \hline
		\hline
	\end{tabular}
	\caption{A benchmark point of our input parameters and observables, where we select it so that $\sqrt{\Delta \chi^2}$ is minimum.}
	\label{bp-tab_nh}
\end{table}
%%%%%%%%%%%%%%%%%%%%%%%%%%%%%%%%%%%%%%%%%%%%%%%%%%%%%%%

\section{Summary and discussion }
\label{sec:conclusion}
We have studied a linear seesaw model with as minimum field content as possible, introducing a modular $S_4$ with the help of $U(1)_{B-L}$ symmetries. Due to rank two neutrino mass matrix, we have had a vanishing neutrino mass eigenvalue. Furthermore, only the normal mass hierarchy of neutrinos is favored through the modular $S_4$ symmetry.
In our numerical $\Delta \chi^2$ analysis, we have found rather sharp prediction on sum of neutrino masses to be around $60$ meV. Imaginary part of $\tau$ is restricted at 1.35-1.5, while real part runs whole the range in the fundamental region. The other remarks are listed below:
 \begin{enumerate}
\item
 $\alpha_{21}$ is within $\{0^\circ - 120^\circ, 240^\circ - 310^\circ \}$,
while $\delta_{CP}$ is within $\{70^\circ - 290^\circ \}$.%
\item 
 $\langle m_{ee} \rangle$ is allowed by $\{0.001 - 0.004\}$ eV.%
 \end{enumerate}
Therefore our model indicates several predictions in neutrino sector that is due to minimal structure with $S_4$ modular symmetry.

%%%%%%%%%%%%%%%%%%%%%%%%%%%%%%%%%%%
\section*{Acknowledgments}
\vspace{0.5cm}
{\it
This research was supported by an appointment to the JRG Program at the APCTP through the Science and Technology Promotion Fund and Lottery Fund of the Korean Government. This was also supported by the Korean Local Governments - Gyeongsangbuk-do Province and Pohang City (H.O.). H. O. is sincerely grateful for the KIAS member.}
%%%%%%%%%%%%%%%%%%%%%%%%%%%%%%%%%%%
%%%%%%%%%%%%%%%%%%%%%%%%%%%%%%%%%%%
%%%%%%%%%%%%%%%%%%%%%%%%%%%%%%%%%%%

\section*{Appendix}

 %%%%%%%%%%%%%%%%%%%%%%%%%%%%%%%%%%%%%%%%%%%%%%%%%%%%%%%%%%%
Here we review some properties of modular $S_4$ symmetry. 
In general, the modular group $\bar\Gamma$ is the group of linear fractional transformation
$\gamma$ acting on the modulus $\tau$ 
which belongs to the upper-half complex plane and transforms as
\begin{equation}\label{eq:tau-SL2Z}
\tau \longrightarrow \gamma\tau= \frac{a\tau + b}{c \tau + d}\ ,~~
{\rm where}~~ a,b,c,d \in \mathbb{Z}~~ {\rm and }~~ ad-bc=1, 
~~ {\rm Im} [\tau]>0 ~.
\end{equation}
This is isomorphic to  $PSL(2,\mathbb{Z})=SL(2,\mathbb{Z})/\{I,-I\}$ transformation.
Then modular transformation is generated by two transformations $S$ and $T$ defined as follows; 
\begin{eqnarray}
S:\tau \longrightarrow -\frac{1}{\tau}\ , \qquad\qquad
T:\tau \longrightarrow \tau + 1\ ,
\end{eqnarray}
and they satisfy the following algebraic relations, 
\begin{equation}
S^2 =\mathbb{I}\ , \qquad (ST)^3 =\mathbb{I}\ .
\end{equation}

Here we introduce the series of groups $\Gamma(N)~ (N=1,2,3,\dots)$ which are defined by
 \begin{align}
 \begin{aligned}
 \Gamma(N)= \left \{ 
 \begin{pmatrix}
 a & b  \\
 c & d  
 \end{pmatrix} \in SL(2,\mathbb{Z})~ ,
 ~~
 \begin{pmatrix}
  a & b  \\
 c & d  
 \end{pmatrix} =
  \begin{pmatrix}
  1 & 0  \\
  0 & 1  
  \end{pmatrix} ~~({\rm mod} N) \right \}
 \end{aligned},
 \end{align}
and we define $\bar\Gamma(2)\equiv \Gamma(2)/\{I,-I\}$ for $N=2$.
Since the element $-I$ does not belong to $\Gamma(N)$
  for $N>2$ case, we have $\bar\Gamma(N)= \Gamma(N)$,
  that are infinite normal subgroup of $\bar \Gamma$ known as principal congruence subgroups.
   We thus obtain finite modular groups as the quotient groups defined by
   $\Gamma_N\equiv \bar \Gamma/\bar \Gamma(N)$.
For these finite groups $\Gamma_N$, $T^N=\mathbb{I}$  is imposed, and
the groups $\Gamma_N$ with $N=2,3,4,5$ are isomorphic to
$S_3$, $A_4$, $S_4$ and $A_5$, respectively \cite{deAdelhartToorop:2011re}.

Modular forms of level $N$ are 
holomorphic functions $f(\tau)$ which are transformed under the action of $\Gamma(N)$ given by
\begin{equation}
f(\gamma\tau)= (c\tau+d)^k f(\tau)~, ~~ \gamma \in \Gamma(N)~ ,
\end{equation}
where $k$ is the so-called as the  modular weight.

Here we discuss the modular symmetric theory framework without imposing supersymmetry explicitly, considering the $S_4$ ($N=4$) modular group. 
Under the modular transformation in Eq.(\ref{eq:tau-SL2Z}), a field $\phi^{(I)}$ is also transformed as 
\begin{equation}
\phi^{(I)} \to (c\tau+d)^{-k_I}\rho^{(I)}(\gamma)\phi^{(I)},
\end{equation}
where  $-k_I$ is the modular weight and $\rho^{(I)}(\gamma)$ denotes an unitary representation matrix of $\gamma\in\Gamma(4)$.
Thus Lagrangian such as Yukawa terms can be invariant if sum of modular weight from fields and modular form in corresponding term is zero (also invariant under $S_4$ and gauge symmetry).

The kinetic terms and quadratic terms of scalar fields can be written by 
\begin{equation}
\sum_I\frac{|\partial_\mu\phi^{(I)}|^2}{(-i\tau+i\bar{\tau})^{k_I}} ~, \quad \sum_I\frac{|\phi^{(I)}|^2}{(-i\tau+i\bar{\tau})^{k_I}} ~,
\label{kinetic}
\end{equation}
which is invariant under the modular transformation and overall factor is eventually absorbed by a field redefinition consistently.
Therefore the Lagrangian associated with these terms should be invariant under the modular symmetry.

The basis of modular forms with weight 2, $Y^{(2)}_{\rm 2} = (y_{1},y_{2})$ and  $Y^{(2)}_{\rm 3'} = (y_{3},y_{4},y_{5})$,  transforming
as a doublet and triplet under $S_4$, are found in terms of the Dedekind eta-function  $\eta(\tau)$ and its derivative \cite{Feruglio:2017spp}:
%%%%%%%%%%%%%%%%%%%%%%%
\begin{eqnarray} 
\label{eq:Y-S3}
y_1(\tau) &=& \frac{i}{8} \left( 8 \frac{\eta'(\tau+\frac12)}{\eta(\tau+\frac12)}  + 32 \frac{\eta'(4\tau)}{\eta(4\tau)}  
- \frac{\eta'(\frac\tau4)}{\eta(\frac\tau4)}- \frac{\eta'(\frac{\tau+1}4)}{\eta(\frac{\tau+1}4)}
- \frac{\eta'(\frac{\tau+2}4)}{\eta(\frac{\tau+2}4)} - \frac{\eta'(\frac{\tau+3}4)}{\eta(\frac{\tau+3}4)}  \right), \nn \\
y_2(\tau) &=& \frac{ i \sqrt3}{8} \left(\frac{\eta'(\frac\tau4)}{\eta(\frac\tau4)} - \frac{\eta'(\frac{\tau+1}4)}{\eta(\frac{\tau+1}4)}
 + \frac{\eta'(\frac{\tau+2}4)}{\eta(\frac{\tau+2}4)} - \frac{\eta'(\frac{\tau+3}4)}{\eta(\frac{\tau+3}4)}  \right)
 \nn \\
%%%
y_3(\tau) &=& i \left(\frac{\eta'(\tau+\frac12)}{\eta(\tau+\frac12)}  - 4 \frac{\eta'(4\tau)}{\eta(4\tau)}  \right),\nn\\
y_4(\tau) &=& \frac{i}{4\sqrt2} \left( - \frac{\eta'(\frac\tau4)}{\eta(\frac\tau4)}+i \frac{\eta'(\frac{\tau+1}4)}{\eta(\frac{\tau+1}4)}
+\frac{\eta'(\frac{\tau+2}4)}{\eta(\frac{\tau+2}4)} - i \frac{\eta'(\frac{\tau+3}4)}{\eta(\frac{\tau+3}4)}  \right), \nn \\
y_5(\tau) &=&  \frac{i}{4\sqrt2} \left( - \frac{\eta'(\frac\tau4)}{\eta(\frac\tau4)} - i \frac{\eta'(\frac{\tau+1}4)}{\eta(\frac{\tau+1}4)}
 + \frac{\eta'(\frac{\tau+2}4)}{\eta(\frac{\tau+2}4)} +i \frac{\eta'(\frac{\tau+3}4)}{\eta(\frac{\tau+3}4)}  \right)
\label{Yi}.
\end{eqnarray}
%%%%%%%%%%%%%%%%%%%%%

$y_i$'s can be expanded in terms of $q$ as follows:
\begin{align}
y_1 &= - 3 \pi \left(\frac{b_1}{8} + 3 b_5\right), \
y_2 = 3 \sqrt{3} \pi b_3, \
y_3 = - \pi \left( -\frac{b_1}{4} + 2 b_5\right), \nn\\
y_4 &= - \pi \sqrt{2} b_2, \
y_5 = - 4\pi \sqrt{2} b_4, 
\end{align}
where $b_i$ are given by 
\begin{align}
b_1 \sim 1, \
b_2 \sim q, \ 
b_3 \sim q^2, \  
b_4 \sim 0, \ 
b_5 \sim 0, 
\end{align}
with $q=\exp( 2 \pi i \tau)$ and $|q| \ll 1$~\cite{Novichkov:2019sqv}.

Then, Yukawas with higher weights are constructed by multiplication rules of $S_4$,
and one finds the following couplings:
%%%%%%%%%%%%%%%%%%%%%%%%%%
\begin{align}
&Y^{(4)}_{\bf2}=
\left[\begin{array}{c}
y_2^2-y_1^2 \\ 
2 y_1 y_2  \\ 
\end{array}\right],\quad
%%%
Y^{(4)}_{\bf3}=
\left[\begin{array}{c}
-2y_2y_3 \\ 
\sqrt3 y_1 y_5+ y_2 y_4  \\ 
\sqrt3 y_1 y_4+ y_2 y_5  \\ 
\end{array}\right],\quad
%%%
Y^{(4)}_{\bf3'}=
\left[\begin{array}{c}
2y_1y_3 \\ 
\sqrt3 y_2 y_5 - y_1 y_4  \\ 
\sqrt3 y_2 y_4- y_1 y_5  \\ 
\end{array}\right]
%,\\
%%%%
%&
%\tilde Y^{(6)}_{\bf3_1}=
%\left[\begin{array}{c}
%y_1^* (y_4^2-y_5^2) \\ 
%y_3 (y_1^* y_5+\sqrt3 y_2^* y_4) \\ 
%-y_3 (y_1^* y_4+\sqrt3 y_2^* y_5) \\ 
%\end{array}\right],\quad
%%
%\tilde Y^{(6)}_{\bf3'_1}=
%(y_1^{*2}+y_2^{*2})
%\left[\begin{array}{c}
%y_3 \\ 
%y_4 \\ 
%y_5 \\ 
%\end{array}\right],\quad
%%
%\tilde Y^{(6)}_{\bf3'_2}=
%\left[\begin{array}{c}
%y_2^* (y_5^2-y_4^2) \\ 
%y_3 (y_2^* y_5-\sqrt3 y_1^* y_4) \\ 
%y_3 (y_2^* y_4-\sqrt3 y_1^* y_5) \\ 
%\end{array}\right]
.
%\nn
\end{align}

%\newref

\end{document}